\documentclass[12pt,preprint]{aastex}

\usepackage{bm}
\usepackage{xcolor}

\newcommand{\vect}[1]{\boldsymbol{\mathbf{#1}}}
\newcommand{\eb}{\begin{equation}}
\newcommand{\ee}{\end{equation}}

\shorttitle{Dislodged AGNs}
\shortauthors{Makarov et al.}
\begin{document}

\title{Astrometric evidence for a population of dislodged AGN} 
\author{Valeri V. Makarov, Julien Frouard, Ciprian T. Berghea}
\affil{United States Naval Observatory, 3450 Massachusetts Ave. NW, Washington, DC 20392-5420, USA}
\email{valeri.makarov@usno.navy.mil}
\author{Armin Rest}
\affil{Space Telescope Science Institute, 3700 San Martin Drive,
Baltimore, MD 21218, USA}
\author{Kenneth C. Chambers, Nicholas Kaiser, Rolf-Peter Kudritzki, Eugene A. Magnier}
\affil{Institute for Astronomy, University of Hawaii at Manoa, Honolulu, HI 96822, USA}

\begin{abstract}
We investigate a sample of 2293 ICRF2 extragalactic radio-loud sources with accurate positions determined by VLBI, mostly active
galactic nuclei (AGN) and quasars,
which are cross-matched with optical sources in the first Gaia release (Gaia DR1). The distribution of offsets between the VLBI
sources and their optical counterparts is strongly non-Gaussian, with powerful wings extending beyond 1 arcsecond. Limiting our
analysis to only high-confidence difference detections, we find (and publish) a list of 188 objects with normalized variances above 12
and offsets below 1 arcsecond. Pan-STARRS stacked and monochromatic images resolve some of these sources indicating the presence of
double sources, confusion sources, or pronounced extended structures. Some 89 high-quality objects, however, do not show any perturbations and
appear to be star-like single sources, yet displaced by multiples of the expected error from the radio-loud AGN. We conclude that
a fraction of luminous AGN (more than 4\%) can be physically dislodged from the optical centers of their parent galaxies.
\end{abstract}

\keywords{astrometry --- reference systems --- quasars: general --- galaxies: nuclei}

\section{Introduction}
\label{Introduction}
The first release of the Gaia mission data (Gaia DR1) includes accurate positions, parallaxes, and proper motions for more than 2 million
Tycho-2 and Hipparcos stars and positions only for the larger Gaia sample of 1.1 billion objects \citep{bro}. In the latter part, the
uncertainty of positions at the mean epoch of J2015 is about 10 mas. Covering the range of $G$ magnitudes between 5 and 20.7, this
sample includes a considerable number of extragalactic sources, including Active Galactic Nuclei (AGN). Radio-loud AGN played an important
role in the production of DR1. The reference frame of this release was adjusted to that of the International Celestial Reference Frame (ICRF2)
\citep{fey} to within 0.1 mas for positions (rotation) and 0.03 mas yr$^{-1}$ for proper motions (spin) \citep{lin}. Optical counterparts 
of 2191 ICRF2 sources (mostly, QSO) were used to this end in a special auxiliary quasar solution combining stellar positions in Hipparcos
at J1991.25 with quasar positions in Gaia at J2015. \citet{mig} used the results of this auxiliary solution to investigate the
astrometric properties of the optical counterparts and the differences between the radio and radio (VLBI) positions. They find no
systematic differences larger than a few tenths of a mas between the Gaia auxiliary solution and confirm that most
of the optical sources are within the expected uncertainties of the VLBI absolute positions.

The presence of large differences in the radio-optical positions of reference AGN of $\sim10$ mas was proposed by \citet{zac} based on 
dedicated CCD observations collected with the 0.9 m telescope, using the UCAC4 astrometric catalog as reference \citep{zac13}. 
\citet{mig} note that most of these
discrepancies may be coming from the larger than expected position errors in the reference catalog, which, in their turn, are probably related to
the considerable sky-correlated pattern in Tycho-2 proper motions \citep{hoga,hogb}. The ultimate source of this very substantial
additional noise in up-to-date astrometric catalogs is probably the Astrographic Catalog 2000 \citep{urb}, which is burdened with
small-scale zonal errors because of the insufficient number of Hipparcos reference stars. More recently, \citet{ber} used the VLBI-measured
radio positions of a larger sample of sources from the OCARS compilation \citep{mal,tit} as hard constraints for a global astrometric
adjustment of the Pan-STARRS catalog and investigated the deviant cases in the process. They determined that a non-negligible fraction
of the VLBI sources ($\sim10$\%) have mismatching optical positions in Pan-STARRS beyond the statistical expectation. A list of sources
was published that should not be used as radio-optical reference frame objects (RORFO) because they exhibit obvious signs of perturbations
on the high-resolution Pan-STARRS maps. The main types of identified problem cases were: extended galaxies with sometimes asymmetric
dust structures, double sources, and microlensing. The relatively low level of astrometric precision in Pan-STARRS (50--70 mas per coordinate)
did not allow
the authors to study the bulk of cases apart from the gross discrepancies.

In this paper, we analyze the best-quality astrometric positions of VLBI sources and their counterparts detected by Gaia. Our approach differs from
the work by \citet{mig} in two respects. We are not attempting to validate Gaia astrometry; instead, we focus on the deviant and special
cases trying to understand the origin of large differences. We use the collected data for optical counterparts from the OCARS selection,
the ICRF2 positions, and the main Gaia
DR1 catalog instead of the special and separate auxiliary solution. Our study continues the study of ``quasometry" special cases
\citep{mak}, aiming at validating and improving the crucially important radio-optical reference frame link. In Section \ref{off.sec},
we briefly describe the sample of cross-matched Gaia-ICRF2-OCARS sources and the relative properties of the radio and optical positions. 
In Section \ref{stat.sec},
our calculation of statistical uncertainties is presented. The smaller selection of statistically significant differences is discussed and
presented in Section \ref{out.sec}, where we employ photometrically calibrated Pan-STARRS images \citep{sch,mag} in the
$grizy$ passbands \citep{ton} to cast light on the nature of some of the outliers. Conclusions are drawn in Section \ref{conc.sec}.

\section{Radio-optical offsets}
\label{off.sec}
We performed a cone search of Gaia DR1 objects within 1$\arcsec$ of each ICRF2 source, which at the time of writing included more than $3\,400$
entries. Counting only the closest counterparts, this resulted in $2293$ tentative matches. The positional matching is straightforward as the extragalactic AGN are believed
to represent a quasi-inertial, non-rotating reference frame, thus no epoch transformations are needed. There is a small number of possible matches with separations greater than 1$\arcsec$, but those are deemed almost certainly confusion sources.

\begin{figure}[htbp]
  \centering
  \includegraphics[angle=0,width=0.7\textwidth]{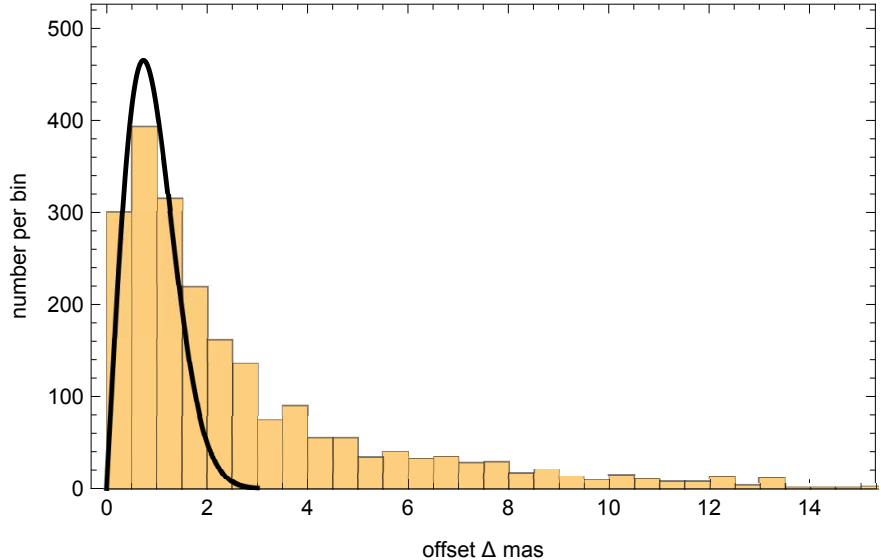}
\caption{Histogram of Gaia--VLBI offsets for the sample of 2293 RORFO. The graph is limited to $\Delta < 15$ mas but
the sample distribution extends to 1 as. The black curve shows the scaled Rayleigh distribution with a scaling
parameter $\sigma=0.73$ mas best fitting the sample distribution.\label{hist.fig}}
\end{figure}

Fig. \ref{hist.fig} shows the distribution of the offsets, or position separations, smaller than 15 mas. The mode of the distribution is
surprisingly small, slightly below 1 mas. At the same time, a long ``exponential" tail is also present stretching all the way to 1$\arcsec$.
The lengths of random 2-vectors with normally distributed coordinates are distributed according to the scaled Rayleigh PDF.
Heuristically, and ignoring at this point the fact that coordinate errors differ for the 2293 vectors, we fit a scaled Rayleigh PDF to describe the
core sample distribution and find a scaling parameter of 0.73 mas (black curve in Fig. \ref{hist.fig}). Approximately  half of the sample
follows this empirical distribution quite well, but the remaining data points cluster at significantly higher values. It appears
that the sample distribution is notably bimodal with a secondary component peaking around 3 mas. However, this visual impression
may be misleading because of the presence of objects with much larger formal errors of coordinates mimicking a secondary population.
A more accurate statistical comparison of found offsets with the expected statistical uncertainties is in order.

\begin{figure}[htbp]
  \centering
  \plottwo{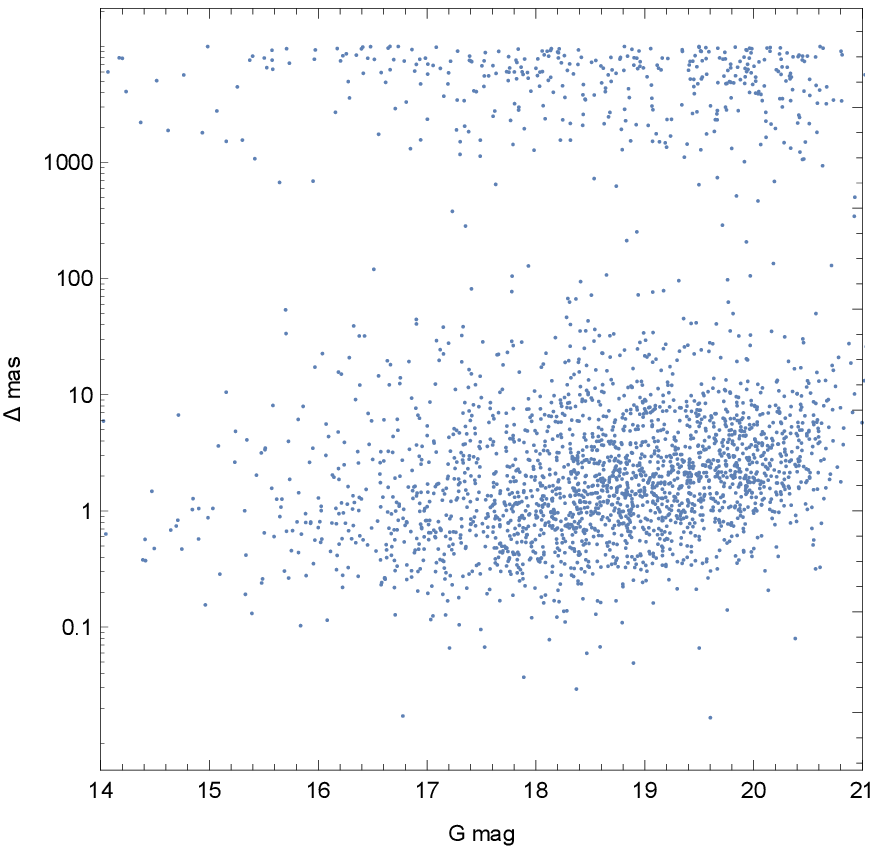}{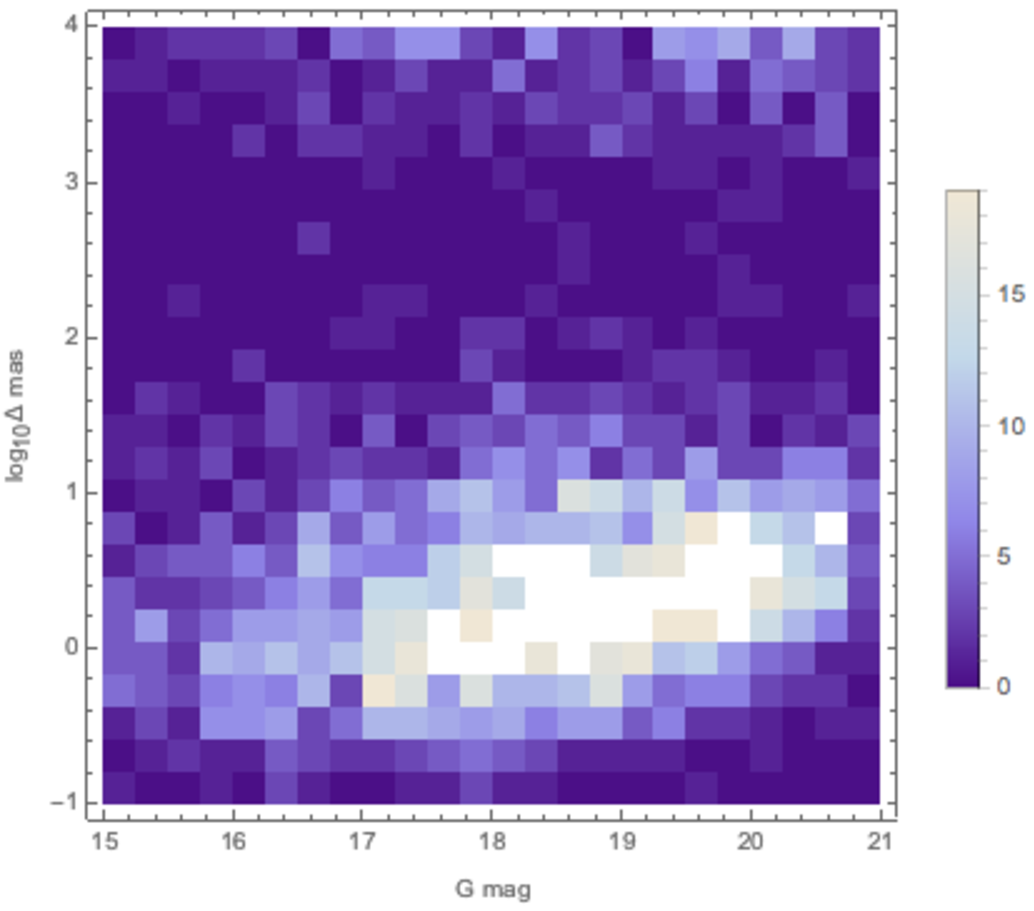}
\caption{Left: Individual position offsets versus G magnitude for cross-matched VLBI sources. Right:
the same distribution but presented as a color-coded number density map. \label{cloud.fig}}
\end{figure}

To make sure we are not missing genuine optical counterparts outside the chosen limit of 1$\arcsec$, we performed a cone search
within 10$\arcsec$ and estimated the number density of matches at different separations. Fig. \ref{cloud.fig}  shows the distribution
of offsets in this wider range as a function of $G$ magnitude as a cloud plot and as a color-coded density map. The so-called error
floor of Gaia DR1 is marked by the slightly curved and densely populated lower envelope, whereas the outliers or perturbed objects
show up as the ``fuzz" at higher offsets $\Delta$ especially evident at magnitudes between 16 and 19. The graphs also show that there is
a gap, or a transition zone, between $\Delta\simeq 100$ and 1000 mas, where relatively few counterparts are found. The cloud above
1000 mas is predominantly confusion sources (unrelated objects happened to be located within the cone). The value of 1000 mas is a conservative
limit leaving most of the genuine counterparts in the sample but also admitting some of the confusion sources, which have to be filtered
out by other means.

\begin{figure}[htbp]
  \centering
  \plottwo{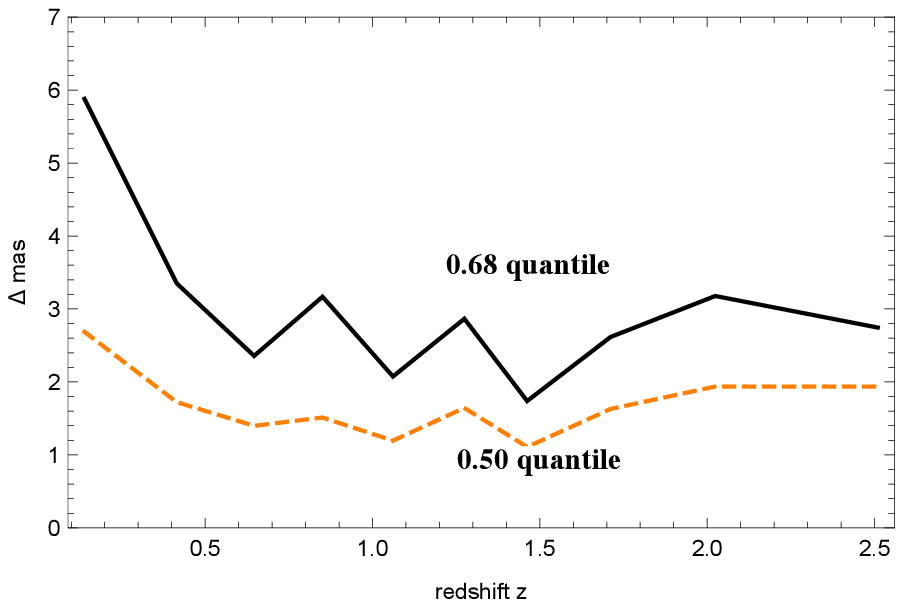}{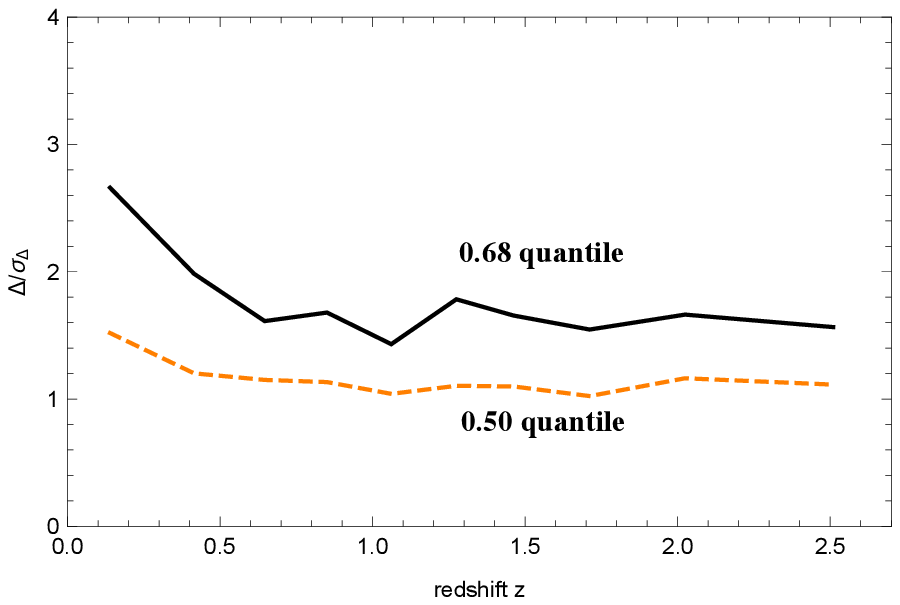}
\caption{Median (0.50 quantile) and 0.68 quantile of  offsets magnitudes (left) and relative offsets $\Delta /\sigma_\Delta$ (right)
versus redshift $z$ for cross-matched VLBI sources in Gaia. \label{quant.fig}}
\end{figure}

To finalize our overview of the general statistical properties of the Gaia-VLBI sample, we present the robust statistics 0.50
and 0.68 quantiles of the frequency distribution of absolute offset magnitudes in mas ($\Delta$, left plot) and relative
offsets ($\Delta/\sigma_\Delta\equiv \sqrt{u}$, right plot) as functions of redshift $z$, Fig. \ref{quant.fig}. Redshifts are available for only 1902
sources in 
our selection. The median values are between 1 and 2 mas for $\Delta$ and slightly above 1 for $\Delta/\sigma_\Delta$, demonstrating
the high quality of astrometry. In fact, the 0.68 quantile for $\Delta/\sigma_\Delta$ is close to 1.5 for $z>0.5$ indicating that
the bulk of offsets at large $z$ have dispersions as expected, since CDF$_{\chi^2_2}(1.5^2)=0.68$. The upturn for $z<0.5$ is quite
noticeable, giving us a clear warning that the astrometric positions of nearby RORFO are perturbed by a detectable effect, likely
of physical nature.

\section{Statistical measures of positional displacement}
\label{stat.sec}
Let a given VLBI source have a radio position defined by a 3-vector ${\bf r}_R$ and an optical counterpart with a Gaia-determined
position ${\bf r}_O$. We can define a pair of basis vectors in the plane tangent to ${\bf r}_R$
\begin{eqnarray}
\vect{\tau}_E &= \langle {\bf p}\times {\bf r}_R \rangle   \nonumber\\
\vect{\tau}_N &= {\bf r}_R \times \vect{\tau}_E
\end{eqnarray}
providing the local East and North directions, respectively, where ${\bf p}=[{\bf 0}\;{\bf 0}\;{\bf 1}]^T$ is the north
pole vector and angular brackets mean normalization to unit length. The two coordinate components of the difference can be computed as
\begin{eqnarray}
d_1 &= (\vect{r}_O-{\bf r}_R) \cdot \vect{\tau}_E  \nonumber\\
d_2 &= (\vect{r}_O-{\bf r}_R) \cdot \vect{\tau}_N
\end{eqnarray}
in the directions of increasing right ascension and declination, respectively. 

Since the position offset is a 2-vector in the tangent plane defined by the unit vectors $\vect{\tau}_E$ and $\vect{\tau}_N$,
and the coordinate measurements are correlated, the statistical significance of coordinate differences $d_1$ and $d_2$ should
include the directional component. The general way to estimate the formal uncertainty in a given direction,
as described by \citet{mig}, is to compute the quadratic form
\eb
u=\vect{d}\, \vect{C}^{-1} \, \vect{d}^T,
\label{u.eq}
\ee
where $\vect{C}$ is the total covariance matrix of the difference vector, $\vect{C}=\vect{C}_O+\vect{C}_R$, and $\vect{d}=[d_1\; d_2]$. 
The quadratic form $u$ is $\chi^2$-distributed with 2 degrees of freedom if the
measurements are normally distributed. For a given position offset, the value of $u$ can be used to estimate the probability
and, thus, the confidence level.

The quadratic form (\ref{u.eq}) is related to the concept of {\it error ellipse} associated with a 2D Gaussian distribution
of a multivariate statistic. The corresponding error ellipse equation is
\eb
u(\vect{d})=1
\label{ell.eq}
\ee
for an arbitrary vector $\vect{d}$ in the tangential plane. Let $\vect{d}=[d_1\; d_2]^T$ be a vector satisfying condition (\ref{ell.eq}),
whose components in polar coordinates are $d_1=\rho\sin \theta$, $d_2=\rho\cos \theta$, where $\theta$ is reckoned  in the
conventional astronomical way from $\vect{\tau}_N$ through $\vect{\tau}_E$. The polar distance $\rho$ can be interpreted
as the standard deviation (1 $\sigma$) of the positional error at the specific position angle $\theta$. The loci of such vectors
describe an ellipse in the tangential plane centered on ${\bf r}_R$ whose size, elongation, and tilt are defined by the
coefficients of the covariance $\vect{C}$. 
There is an excess of values close to
$+1$ or $-1$ in the overall distribution of coordinate correlation coefficients specified in the Gaia DR1 catalog. This
arises from a non-uniform distribution of scan angles, quite frequent for this limited span of the mission. These large correlations
and unequal variances of coordinates imply very elongated error ellipses, where the position uncertainty strongly varies with
the position angle. 

In the following, we will focus on extreme statistical outliers. The $u$-statistic allows us to assign specific confidence to
each observed offset. The null hypothesis in this case is that a given observed offset  $\vect{d}$ is a random outcome of the
expected 2D Gaussian PDF defined by the covariance $\vect{C}$. The confidence level that the null hypothesis is rejected is
then simply CDF$_{\chi^2_2}(u)$. For the selection investigated in this paper, we set a threshold value of $u_{\rm lim}=12$,
which corresponds to a confidence of 0.99752. The $p$-value (i.e., the probability of rejecting the null hypothesis while being
true) is 0.0025. For our sample of 2293 RORFO, this translates to an expectation of 5.7 false positives. The number of outliers we
find with this criterion is 188.

\section{Objects with large radio-optical offsets}
\label{out.sec}
In Section \ref{off.sec}, we found that the core distribution of Gaia-VLBI position offsets has the expected dispersion
specified by the formal measurement errors quite well. The robust statistics shows only a small ($\sim10-20$\%) excess
in normalized errors. The presence of a smaller fraction of extreme outliers, on the other hand, is betrayed by the
upturn of offset magnitudes at small redshifts (Fig. \ref{quant.fig}) and by an extended wing in the sample distribution
on Fig. \ref{hist.fig}. 

Our entire statistical selection of extreme outliers is published online. Table~1 provides a small cut-out of the file.
Apart from the information carried over from ICRF2 (columns 2 -- 5), OCARS (columns 7 -- 8) and Gaia (column 6) for users' convenience,
we add the following data. Column 1 gives the offset magnitudes in mas. 
Column 9 contains the total number of Gaia sources found within $10\arcmin$ of a given
VLBI position. This number is used to compute for each entry the probability of finding an unrelated confusion source
(interloper) within $1\arcsec$ of this position, given in column 10 in percent. Even though the interloper probability
is non-negligible for some of the sources, the overall contamination is expected to be small. The estimated number of chance alignments
for the entire sample of 2293 sources is 9.1. Even in case all these interlopers happened to be selected, they may only account
for a small fraction of detected offsets.

\begin{deluxetable}{lrrrrrrrrrl}
\tablecaption{Objects with large radio-optical offsets \label{obj.tab}}
\tablewidth{0pt}
\tablehead{
\multicolumn{1}{r}{(1)$^\dag$}  &
\multicolumn{1}{r}{(2)}  &
\multicolumn{1}{r}{(3)}  &
\multicolumn{1}{r}{(4)}  &
\multicolumn{1}{r}{(5)}  &
\multicolumn{1}{r}{(6)}  &
\multicolumn{1}{r}{(7)}  &
\multicolumn{1}{r}{(8)}  &
\multicolumn{1}{r}{(9)}  &
\multicolumn{1}{r}{(10)}  &
\multicolumn{1}{c}{(11)}   \\
}

\rotate \tabletypesize{\scriptsize} \startdata
6.61 & J125759.0-315516 & 194.4960867 & $-31.92134769$ & N & 17.410 & 1.924 & AQ & 843 & 0 & \\
6.53 & J074125.7+270645 & 115.3572201 & 27.11260883 & V & 19.480 & 0.7721 & AQ & 673 & 0 & galaxy\\
6.51 & J013957.3+013146 & 24.98877424 & 1.529482913 & N & 18.141 & 0.2617 & AS & 201 & 0 & double, red comp. 1.5" 300 deg\\
6.50 & J080757.5+043234 & 121.989744 & 4.542925278 & D & 18.189 & 2.877 & AQ & 1079 & 0 & *\\
6.22 & J074447.2+212000 & 116.1969863 & 21.33345193 & N & 19.787 & 2.5047 & AQ & 815 & 0 & *\\
\enddata
\tablenotetext{\dag}{Columns:\newline
(1) Absolute difference in position between Gaia and VLBI, mas;
(2) ICRF2 name; (3) ICRF2 right ascension, deg; (4) ICRF2 declination, deg; (5) ICRF2 type: D for defining, V for VCS, N for non-VCS; (6) 
G magnitude; (7) redshift from OCARS; (8) morphological type from OCARS;  (9) number of Gaia sources within $10\arcmin$; (10) probability of chance
neighbor within $1\arcsec$, \%; (11) notes (an asterisk * stands for a star-like image, galaxy for obviously extended objects).}
\end{deluxetable}

\begin{figure}[htbp]
  \centering
  \includegraphics[angle=0,width=0.3\textwidth]{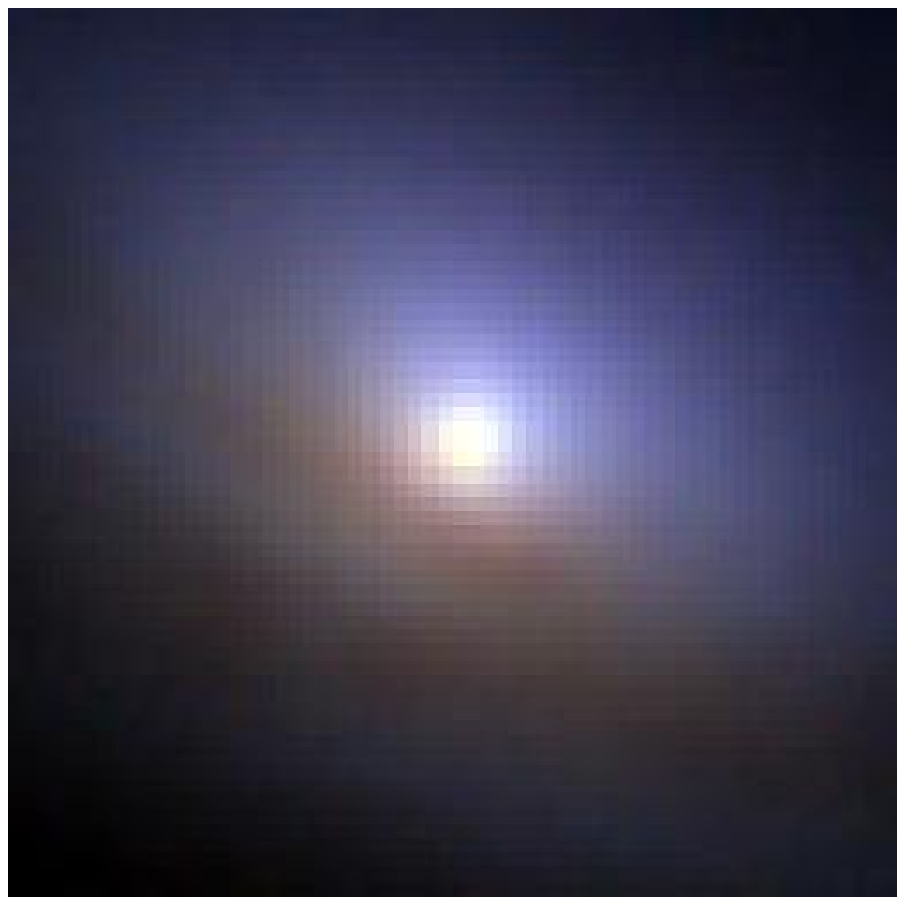}
  \includegraphics[angle=0,width=0.3\textwidth]{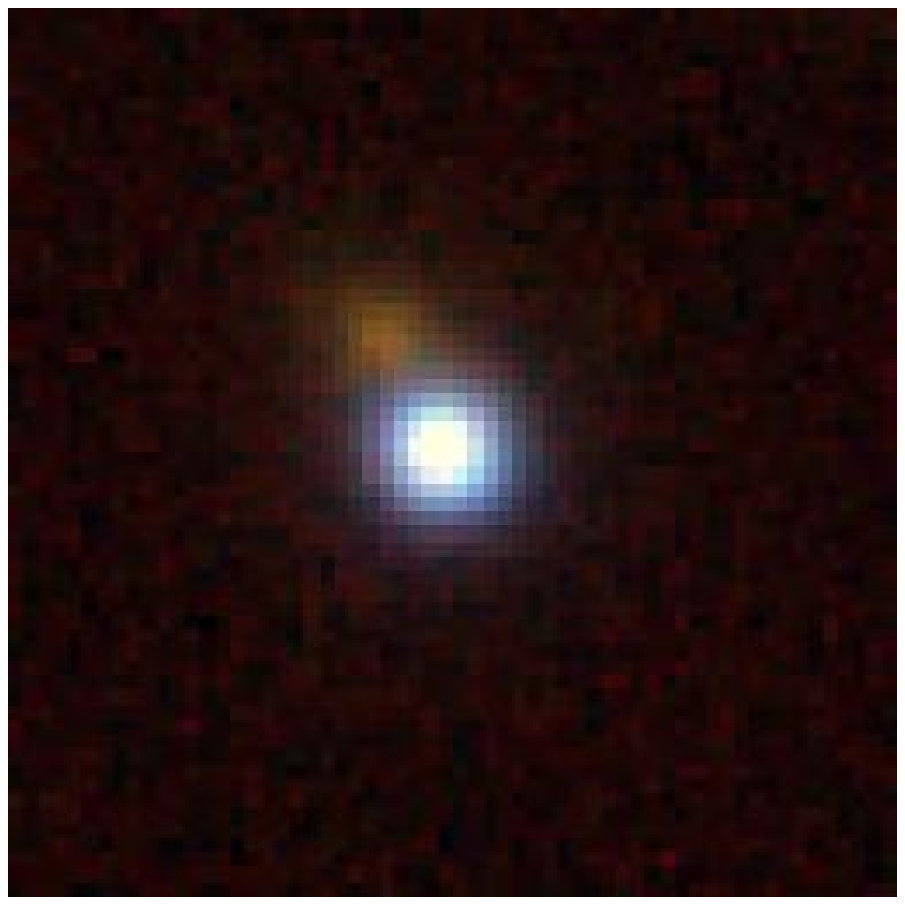}
  \includegraphics[angle=0,width=0.3\textwidth]{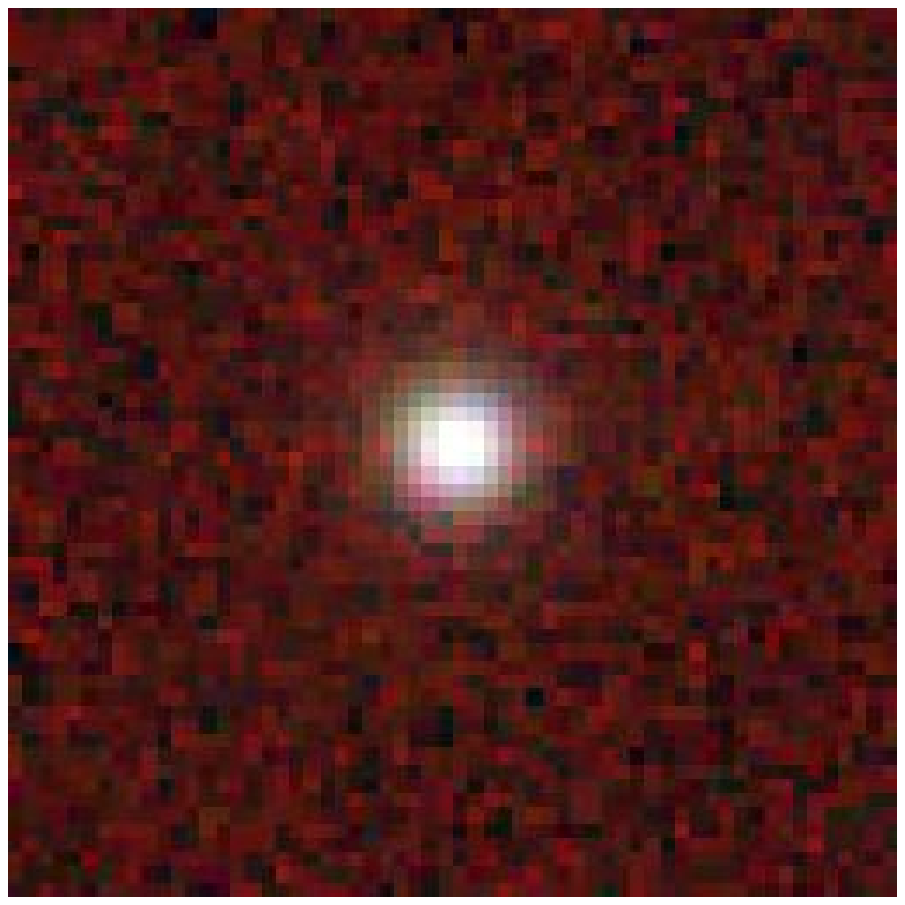}
\caption{Pan-STARRS images of three representative types of optical counterparts of VLBI-measured radio-loud AGN. Each image
is $15\arcsec$ on a side, with North up and East to the left. The matched radio source is
exactly at the center of each image. Left: a galaxy with a dust structure.  Middle: a resolved double source
of different color. Right: a star-like single source. \label{map.fig}}
\end{figure}

Some of the AGN with discrepant radio-optical positions are residing in the cores of luminous galaxies. Such extended
objects probably explain why the frequency of outliers is higher at small redshifts (Fig. \ref{quant.fig}). This was our
motivation to visually inspect colored stacked and monochromatic Pan-STARRS images in the pre-release version available at
the STScI. Pan-STARRS is a multi-passband panoramic survey of the northern 3/4 of the sky (above Dec$\simeq -30\degr$),
so only three quarters of the sample can be reviewed this way. The quality of Pan-STARRS images is superior to the previous 
surveys reaching $1\arcsec$ resolution and better quite routinely. This inspection of images revealed three main
morphological types of RORFO with discrepant positions. First, as we expected, a significant fraction of the sample corresponds
to extended objects, most likely to be the host galaxies (see the example in Fig. \ref{map.fig}, left). These are marked in the last 
column (11) as ``galaxy". OCARS sometimes includes specific morphological classification of the optical counterpart, reproduced
in column (8). We find a clear correlation between the types G (galaxy) and AS (seyfert) and our extended
sources resolved by Pan-STARRS. Some of the galaxies have clearly visible dust structures obscuring one side of the image
more than the other and thus generating a bias of the optical photocenter. We note these objects as ``galaxies with dust".
We find 35 extended objects in the sample. Taking into account that one quarter of the sample is unavailable for
inspection, the rate of galaxies is 25\%. The large offsets for this type are not surprising, because of possibly asymmetric
profiles and inevitable errors in Gaia astrometry, which was not adjusted to non-pointlike sources. A single template line spread function
was determined in DR1 for each CCD and each gate \citep{fab} which precluded accurate astrometry of galaxies. In fact, it is more
surprising that such extended images sneaked into Gaia DR1 at all, as many other OCARS galaxies were apparently discarded by the
pipeline.

The second morphological type is a double source, which we find in 25 cases (see example in Fig. \ref{map.fig}, middle),
with an estimated rate of 18\%. We note visually estimated angular resolution
and position angle of such double sources. The separation ranges between $<1\arcsec$ and $3\arcsec$, approximately. Separations
larger than $3\arcsec$ were seen but not recorded because those can hardly perturb Gaia astrometry \citep{fab}. In principle,
all these occurrences can be attributed to chance alignment; however, some fraction of them can be physical double galaxies or
mergers because confusion sources would not generate an optical pair for those ICRF2 counterparts that are simply too faint
or too perturbed to be detected by Gaia. At any rate, there seems to be a clear reason for the optical positions to be
perturbed in this case. A few cases of possibly multiple, tightly packed images are also recorded, which may be microlenses.

The third category, and the most interesting in the context of this study, are star-like objects that do not show any obvious
perturbation in Pan-STARRS images. They look single and compact and are marked with ``*" in column 11. An example of such an object is
shown in Fig. \ref{map.fig}, right. It is certainly possible that some of them are unresolved extended asymmetric objects or doubles.
The probabilities of chance alignment range between 0 and a few percent. We count 89 of these objects in total. Those
of them that have a definite morphological class in OCARS are mostly AQ (quasars), with a smaller percentage of AL (lacertae). 
The sample of 89 star-like sources with large optical--radio position offsets requires further scrutiny and follow-up observations.

\section{Conclusions}
\label{conc.sec}
We have determined that the core distribution of Gaia$-$VLBI position offsets obeys the expected PDF, but a significant
fraction of the matches has gross differences in positions that can not be explained by the estimated random or systematic
errors. Focusing only on high-confidence cases of discrepancy, we selected 188 objects with offsets up to 1000 mas.
About 25\% of the latter selection are resolved in Pan-STARRS images as extended objects (galaxies), sometimes accompanied by
asymmetric dust structures, which can shift the optical photocenters. Another 18\% of the sample of outliers are resolved
as double (or multiple) sources, probably chance alignment of unrelated objects \citep{ass}, and possibly some genuine double
galaxies. The remaining 57\% of objects available in the Pan-STARRS survey appear to be single and star-like, without any
obvious problems or perturbations. 
We propose that some of the surviving 89 objects may represent a population of physically dislodged AGN
which are not residing in the centers of the host galaxies.

The robust statistics of dispersion presented in Fig. \ref{quant.fig} suggest that the optical positions of RORFO
are dispersed in equal measure at any redshift except for $z<0.5$ where a larger scatter is present. The simplest and the likeliest
explanation is that the remaining nearby galaxies with AGN that somehow propagated into the Gaia DR1 catalog possess a larger
degree of asymmetries, probably related to dust structures and merger remnants. The same effect should be present in more
distant host galaxies, but reduced proportionally to distance. The flatness of the dispersion curves suggests that
the extended components of images become insignificant at a certain typical distance. \citet{ham} proposed a ``fundamental plane"
relation based on collected observations of 70 AGN with $0.06 \leq z \leq 0.46$, according to which the magnitude
difference in $V$ between the host galaxy and the nucleus is larger for luminous nuclei. The statistical equality is achieved
at $M_V({\rm nuc})=-22.8$, but already at $M_V({\rm nuc})=-25.7$, the host galaxy is typically 2 mag fainter than the nucleus.
At higher redshifts, only intrinsically bright nuclei can be detected by Gaia. Therefore, we should expect even smaller
dispersions of position offsets at $z>1$. This is obviously not the case. Either the impact of host galaxy asymmetries is
completely switched off at approximately $z=0.5$, or the astrometric precision of Gaia DR1 measurements is insufficient to detect it.

The sample of double sources can be explained as chance alignments (e.g., with foreground galactic stars) if nearly all of them
ended up in our selection of extreme outliers. This possibility seems more likely with the limited data in hand, because
many of the doubles have distinctly different colors in Pan-STARRS stacked images. We note that, due to the proper motion of galactic stars,
the optical photocenters of such blended images should be variable in time. A smaller fraction of this sample may
still represent the interesting class of double galaxies. Spectroscopic observations are needed to find these. At separations
$1\arcsec$ -- $3\arcsec$ recorded in our paper, many of them are not interacting or merging pairs. 

In a similar study, \citet{oro} used a combination of ICRF2 and SDSS DR9, resulting in a smaller initial sample of cross-matched 
objects. The astrometric accuracy of SDSS data is much poorer ($\sim 55$ mas per coordinate) than that of Gaia DR1, but the
authors were able to statistically estimate that about 4\% of their sample display significant differences in excess of $3\sigma$.
They proposed that the cases of large positional offsets represent yet unknown strong gravitational lenses or binary AGN candidates.
To verify if some of the double sources found in our paper are gravitational lenses, accurate spectroscopy and epoch photometry
follow-up is required.

About 88\% of our sample of significant displacements have magnitudes smaller than 100 mas, where the diagnostic power of
Pan-STARRS images becomes limited. Therefore, a fraction of of the sample is bound to include yet unresolved extended galaxies
and structures, as well as double nuclei. This issue can be resolved with optical imaging of much higher resolution, e.g., with the HST.
Kiloparsec-sized, unresolved jet structures of considerable optical brightness represent another intriguing possibility (although no such
structures were found in nearby galaxies). We performed a limited study of this possibility using the collection of radio-detected jets
in \citep{moo}. Using 57 objects in common with our initial ICRF2 sample, we investigated the differences in position angles between the
radio-loud jets and Gaia-ICRF2 offsets. The null hypothesis that these differences (modulo 180$\degr$) are uniformly distributed obtains
large $p$-values, e.g., 0.41 with the Pearson $\chi^2$ test, and 0.50 with the Watson $U^2$ test. Thus, we found no support for the
jet hypothesis within this limited sample.

The most interesting result of this study is a sample of star-like, single objects that are likely to include
a population of AGN dislodged from their host galaxies' centers. Specific cases of this scenario have been discussed in
the literature. \citet{hof} proposed that a binary supermassive black hole can dynamically interact with a quasar during
a merger event, imparting the latter with a sufficient kick to drive it outside of the stellar spheroid. The quasar
continues to be fueled by the gas kicked out along with it, but for a limited duration of time. Gravitational wave
recoil and slingshot recoil are two competing theories to explain a few known double nuclei in relatively nearby
galaxies \citep[e.g.,][]{civ}. \citet{kom} propose that kicked-out SMBH can obtain velocities of thousands km s$^{-1}$,
which may be enough to completely dislodge them from the host galaxies. The merger process is thought to be much more effective at
earlier epochs ($z\sim 2$) \citep{spr}, where double nuclei should be rather common, but they are harder to detect at these
distances. Very few OCARS sources have been observed with high-resolution imaging instruments of the HST, while the resolution
of ground-based telescopes and X-ray detectors is insufficient to explore the sub-1$\arcsec$ domain. Gaia astrometry combined
with VLBI astrometry provides a much more powerful method of detection of dislodged AGN with sensitivity 100 times better
than the regular imaging surveys. Even higher capabilities will be achieved with the future Gaia releases. The sample of
apparently single and unperturbed sources (mostly quasars) detected astrometrically in this paper includes prime candidates
to new objects of this kind in a significantly wider range of redshifts.

\acknowledgments
This work has made use of data from the European Space Agency (ESA)
mission {\it Gaia} (\url{http://www.cosmos.esa.int/gaia}), processed by
the {\it Gaia} Data Processing and Analysis Consortium (DPAC,
\url{http://www.cosmos.esa.int/web/gaia/dpac/consortium}). Funding
for the DPAC has been provided by national institutions, in particular
the institutions participating in the {\it Gaia} Multilateral Agreement. The Pan-STARRS1 Surveys (PS1) have been made possible through contributions of the Institute for Astronomy, the University of Hawaii, the Pan-STARRS Project Office, the Max-Planck Society and its participating institutes, the Max Planck Institute for Astronomy, Heidelberg and the Max Planck Institute for Extraterrestrial Physics, Garching, The Johns Hopkins University, Durham University, the University of Edinburgh, Queen's University Belfast, the Harvard-Smithsonian Center for Astrophysics, the Las Cumbres Observatory Global Telescope Network Incorporated, the National Central University of Taiwan, the Space Telescope Science Institute, the National Aeronautics and Space Administration under Grant No. NNX08AR22G issued through the Planetary Science Division of the NASA Science Mission Directorate, the National Science Foundation under Grant No. AST-1238877, the University of Maryland, and Eotvos Lorand University (ELTE) and the Los Alamos National Laboratory.

\end{document}